\documentclass[conference]{IEEEtran}
\usepackage{inputenc}
\usepackage{fontenc}
\usepackage{mathtext}
\usepackage{textcase}
\usepackage{bm}
\usepackage{amsfonts,amssymb}
\usepackage{subfigure}
\usepackage{mathrsfs}
\usepackage{color}
\usepackage[ruled]{algorithm2e}
\usepackage{textcomp}
\usepackage{multirow}

\usepackage{ifpdf}
 \ifpdf
 \else
 \fi

\usepackage{cite}

\ifCLASSINFOpdf
  \usepackage[pdftex]{graphicx}
\else
   \usepackage[dvips]{graphicx}
   
\fi
\usepackage{epstopdf}
\usepackage{caption}
\usepackage{amsmath}
\usepackage{algorithmic}

\usepackage{array}
\begin{document}
%
\title{Rate-Splitting Multiple Access for Multi-Antenna Joint Communication and Radar Transmissions}
%
%
%

\author{\IEEEauthorblockN{Chengcheng Xu\IEEEauthorrefmark{1},
Bruno Clerckx\IEEEauthorrefmark{2},
Shiwa Chen\IEEEauthorrefmark{1}, 
Yijie Mao\IEEEauthorrefmark{2} and
Jianyun Zhang\IEEEauthorrefmark{1}}
\IEEEauthorblockA{\IEEEauthorrefmark{1}College of Electronic Engineering\\
National University of Defense Technology,
Hefei, China\\ Email: \{xuchengcheng17,chenshiwa17,zhangjianyun17\}@nudt.edu.cn}
\IEEEauthorblockA{\IEEEauthorrefmark{2}Department of Electrical and Electronic Engineering\\
	Imperial College London,
	London, U.K.\\
Email: \{b.clerckx, y.mao16\}@imperial.ac.uk}
}
\maketitle

\begin{abstract}
Rate-Splitting Multiple Access (RSMA), relying on multi-antenna Rate-Splitting (RS) techniques, has emerged as a powerful strategy for multi-user multi-antenna systems. In this paper, RSMA is introduced as a unified multiple access for multi-antenna radar-communication (RadCom) system, where the base station has a dual communication and radar capability to simultaneously communicate with downlink users and probe detection signals to azimuth angles of interests. Using RS, messages are split into common and private parts, then encoded into common and private streams before being precoded and transmitted. We design the message split and the precoders for this RadCom system such that the Weighted Sum Rate (WSR) is maximized and the transmit beampattern is approximated to the desired radar beampattern under an average transmit power constraint at each antenna. We then propose a framework based on Alternating Direction Method of Multipliers (ADMM) to solve the complicated non-convex optimization problem. Results highlight the benefits of RSMA to unify RadCom transmissions and to manage the interference among radar and communications, over the conventional Space-Division Multiple Access (SDMA) technique.
\end{abstract}
%

%
\IEEEpeerreviewmaketitle

\section{Introduction}
The 4th and 5th generation wireless communication systems are competing with long-range radar applications in the S-band (2-4GHz) and C-band (4-8GHz), which will possibly result in severe spectrum congestion and hamper the higher data rate requirements for the increasing demand in future wireless communication.\cite{6967722}. Though efforts for new spectrum management regulations and policies are needed, a longer term solution is to enable communication and radar spectrum sharing (CRSS). There are two main research topics in the field of CRSS: 1) coexistence of existing radar and communication devices, 2) co-design for dual-function systems. \par 
For coexistence of existing radar and communication devices, research focuses on designing high-quality wideband radar waveforms that achieve spectrum nulls on communication frequency bands\cite{Aubry2016,8528529}, as well as on jointly designing communication precoders and slow-time radar waveforms to meet radar Signal-to-Interference-plus-Noise (SINR) and communication rate requirements\cite{Zheng2018}. All the aforementioned works are limited to single-antenna radar systems. As multi-antenna processing can greatly improve radar performance\cite{li2007mimo}, research has been devoted to the coexistence of existing Multiple-Input Multiple-Output (MIMO) communication systems and MIMO radar systems\cite{qian2018joint}. However, given the existing infrastructure, a coexistence approach manages interference between radar and communication as much as it can, while a joint design approach makes the best use of the spectrum for the dual purpose of detecting and communicating. As a consequence, a joint radar and communication system design approach would outperform a coexistence approach. Early studies \cite{saddik2007ultra,sturm2011waveform} consider single-antenna dual-function platforms without utilizing multi-antenna processing. In \cite{hassanien2015dual,hassanien2016phase} the information stream is embedded in radar pulses via a multi-antenna platform, leading to the restriction that the rate is limited by the Pulse Repetition Frequency (PRF), which is far from satisfactory for communication requirements. To overcome this restriction, \cite{liu2018mu,liu2018toward} propose a joint multi-antenna radar-communication (RadCom) system that simultaneously transmits probing signals to radar targets and serves multiple downlink users. The precoders are designed to form a desired radar beampattern and meet the SINR requirements for communication users. \par 
The main problem in the multi-antenna RadCom system is how to efficiently manage the interference among radar and communication users. In the past few years, a powerful and versatile framework of multi-antenna non-orthogonal transmission and interference management strategy based on Rate-Splitting (RS) has emerged \cite{7470942,joudeh2016sum,dai2016rate,joudeh2017rate,mao2019rate,mao2018rate}. The flexibility of RS comes in the potential to partially decode interference and partially treat it as noise, through message splitting and the creation of common and private streams. 
As a consequence, rate-splitting multiple access (RSMA) brings rate benefits over space-division multiple access (SDMA) and non-orthogonal multiple access (NOMA)\cite{mao2018rate}.\par 
In this paper, we introduce a novel way to design RadCom and show the potential and versatility of RS in multi-antenna RadCom system to manage jointly multi-user interference and communication-radar interference. Uniquely, the benefit of RS originates from the presence of the common stream that is not only used to manage interference between communication users but also better approximate the desired radar beampattern. Specifically, we are motivated to design a multi-antenna RadCom system that functions as both a communication BS and a collocated MIMO radar, so as to maximize the Weighted Sum Rate (WSR) of users and to detect targets following a desired radar beampattern. Firstly, we build the RSMA-based multi-antenna RadCom system model, which is the first work that combines RS and RadCom system design to the best of our knowledge. Secondly, we formulate the problem of maximizing the WSR of all users and approximating the transmit beampattern to the desired one with an average transmit power constraint at each antenna. Thirdly, to solve the complicated non-convex problem, we propose an ADMM-based iterative method. Finally, we compare the performance of our RSMA-based RadCom with the SDMA-based RadCom system. Numerical results show that the proposed RSMA-based RadCom system enables a better tradeoff between WSR and beampattern approximation compared with the SDMA-based RadCom system.

%
%
%
%

\section{System Model and Problem Formulation}
\begin{figure}
	\centering
	\includegraphics[width=6cm]{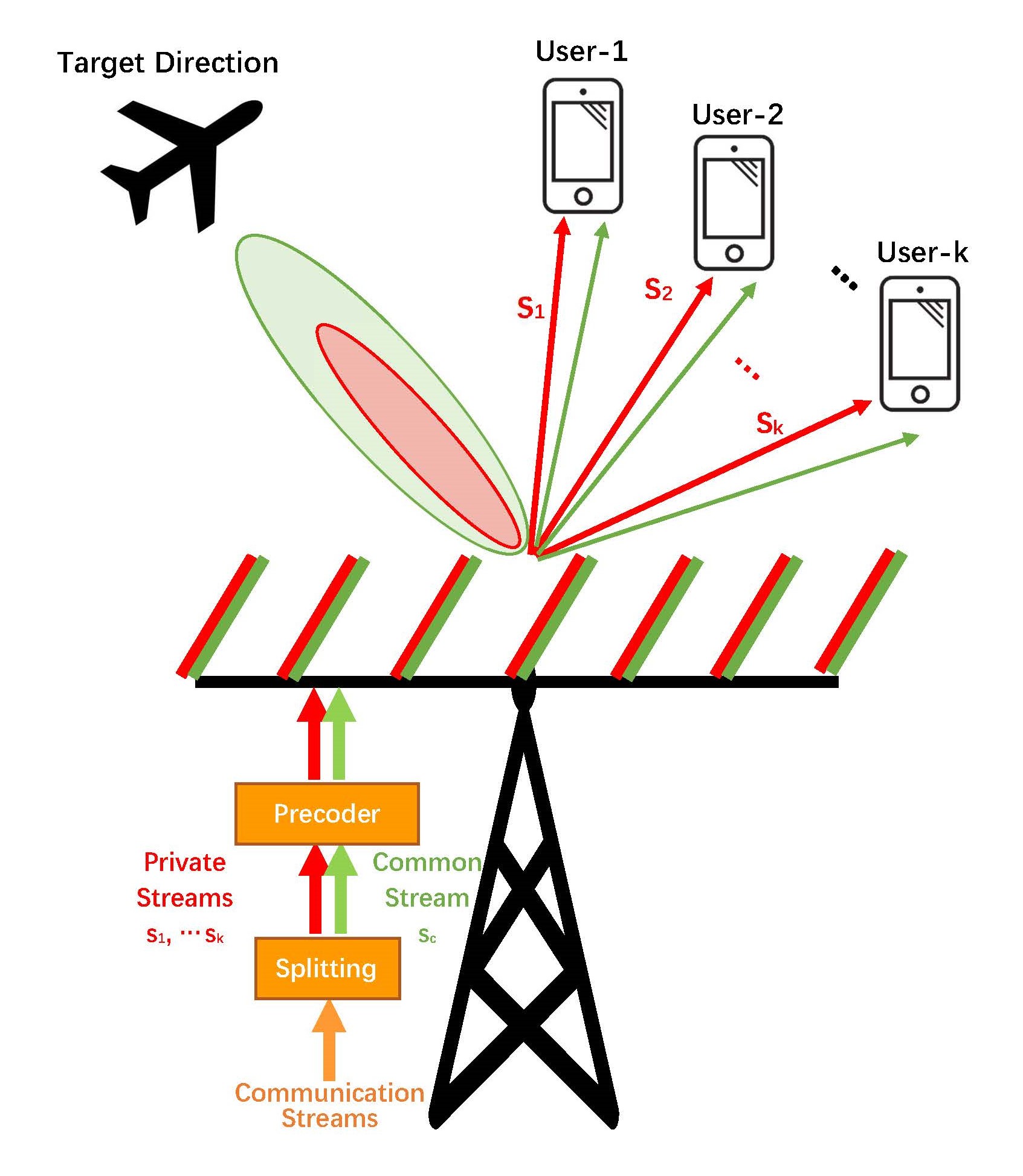}
\caption{Schematic diagram for the proposed multi-antenna RadCom system.}\label{scheme}
\end{figure}

In this work, we consider a downlink multi-antenna RadCom system, which is equipped with a uniform linear array (ULA) of $N_t$ antennas and serves $K$ single-antenna communication users and one radar target. The communication users are indexed by $\mathcal{K} =\{1,\dots, K\}$. The schematic diagram of the proposed system is shown in Fig. \ref{scheme}.\par 

\subsection{Rate-Splitting Preliminaries}
Rate-splitting is a promising transmission technique to tackle numerous
problems faced by modern MIMO wireless networks\cite{7470942,mao2018rate}. The message $W_k$ of the $k$th user is split into a common part $W_{c,k}$ and a private part $W_{p,k}$, $\forall k\in \mathcal{K}$. The common parts of all users $\{W_{c,1}, \dots, W_{c,K}\}$ are jointly encoded into the common stream $s_c$, while the private parts $\{W_{p,1}, \dots, W_{p,K}\}$ are respectively encoded into private streams $\{s_{1}, \dots, s_{K}\}$. Then the data stream vector ${\bf{s}}=[s_c,s_1,\dots,s_k]^T \in \mathbb{C}^{K+1}$ is linearly precoded using the precoder ${\bf P}=[{\bf{p}}_c,{\bf p}_1,\dots,{\bf{p}}_K]$, where ${\bf{p}}_c\in \mathbb{C}^{N_t\times1}$ is the precoder of the common stream.\par 
\subsection{MIMO Radar Beampattern Design}
MIMO radar can transmit multiple probing signals that may be chosen freely via its antennas. Such waveform diversity enables superior capabilities compared with standard phased-array radar\cite{li2007mimo}. To achieve higher signal-to-noise ratio (SNR) for target tracking or narrow-beam scanning scenarios, and to avoid wasting probing power on either jammer locations or locations of uninteresting targets, a specially desired rather than a nominally omnidirectional MIMO radar beampattern should be achieved via correlated waveform design\cite{li2007range,fuhrmann2008transmit,stoica2007probing}. According to \cite{ fuhrmann2008transmit,stoica2007probing}, the beampattern design problem can be formulated in a more explicit way
\begin{equation}
\begin{split}
\quad\quad\quad\min\limits_{\alpha,{\bf R}} \quad &\sum_{m=1}^{M}\left|\alpha P_d(\theta_m)-{\bf a}^H(\theta_m){\bf R}{\bf a}(\theta_m)\right|^2 \\
s.t.\quad&{\text{diag}}({\bf R})=\frac{P_0{\bf 1}}{N_t}\\
&{\bf R}\in \mathbb{S}^{N_t}_+\\
&\alpha>0
\end{split}\label{beamformingprob}
\end{equation}
where $\theta_m$ is the $m$th azimuth angle grid among all $M$ grids, $P_d(\theta_m)$ is the desired beampattern level at $\theta_m$, $\bf R$ is the covariance matrix of transmit waveforms , ${\bf a}(\theta_m)=[1,e^{j2\pi\delta sin(\theta_m)},\dots,e^{j2\pi(N_t-1)\delta sin(\theta_m)}]^T\in\mathbb{C}^{N_t\times 1}$ is the transmit steering vector, $\delta$ is the normalized distance (relative to wavelength) between adjacent array elements and $\text{diag}(\cdot)$ refers to the vector constructed from the diagonal entries of the matrix. In \eqref{beamformingprob}, the objective function is to approximate the beampattern to the desired one via least squares, the first constraint ensures that every radar transmit antenna has the same average power with total power budget $P_0$, and the second constraint ensures that $\bf R$ is a semi-positive definite matrix.\par 
\subsection{Joint RadCom Transmission System Design}
Our multi-antenna RadCom system works as a RS-assisted BS, and simultaneously as a collocated MIMO radar, i.e., a mono-static radar. The transmit signal is formulated as  
\begin{equation}
{\bf x}[l]={\bf p}_c{s}_c[l]+\sum_{k\in \mathcal{K}}{\bf p}_k{s}_k[l].\label{transig}
\end{equation}
where $s_c[l]$ is the common stream  in the RS strategy, $s_k[l]$ is the private stream for the $k$th user, and $l$ is the time index. We only adopt precoded information streams as the transmit signal, but \eqref{transig} meets the MIMO radar transmit signal model and thus are adequate for MIMO radar detecting according to \cite{onTransMIMO}. Based on this transmit signal model, \cite{onTransMIMO} derives that improving the SNR of the matched-filter output at the MIMO radar receivers, which is a key metric for radar detection, is equivalent to designing a desired radar transmit beampattern. As a result, we focus on increasing communication WSR and designing desired radar transmit beampattern in this paper.\par     
Based on \eqref{transig}, the received signal at the $k$th user is
\begin{equation}
\begin{aligned}
y_k[l]=&{\bf h}_k^H{\bf x}[l]+n_k[l], \forall k\in \mathcal{K}\\
=&{\bf h}_k^H{\bf p}_c{s}_c[l]+{\bf h}_k^H{\bf p}_k{s}_k[l]+{\bf h}_k^H\sum_{j\in \mathcal{K},j\neq k}{\bf p}_j{s}_j[l]+n_k[l]
\end{aligned}
\end{equation} 
where ${\bf h}_k \in \mathbb{C}^{N_t}\times 1$ is the channel vector between the RadCom system and the $k$th user. It is assumed to be perfectly known at both the transmitter and the receivers. The received noise $n_k[l]$ at the $k$th user is modelled as a complex Gaussian random variable with zero mean and variance $\sigma _{n,k}^2$. Without loss of generality, we assume the noise variances all equal to 1, i.e., $\sigma _{n,k}^2=1, \forall k \in \mathcal{K}$. \par 
Following the decoding order in the literature of RS\cite{joudeh2016sum}, each user first decodes the common stream by treating all private streams as interference. Therefore, the SINR for decoding ${s}_c$ at the $k$th user is
\begin{equation}
\gamma_{c,k}({\bf P})=\frac{\left| {\bf h}_k^H{\bf p}_c\right| ^2}{\sum_{j\in \mathcal{K}}\left| {\bf h}_k^H{\bf p}_j\right|^2+1},\forall k \in \mathcal{K}.\label{CO sinrck}
\end{equation}\par 
After successfully decoding ${s}_c$ and subtracting it from $y_k$, user-$k$ decodes the intended private stream ${s}_k$ by treating other private streams as interference. The SINR of decoding ${s}_k$ at user-$k$ is
\begin{equation}
\gamma_k({\bf P})=\frac{\left| {\bf h}_k^H{\bf p}_k\right| ^2}{\sum_{j\in \mathcal{K},j\neq k}\left| {\bf h}_k^H{\bf p}_j\right|^2+1},\forall k \in \mathcal{K}.\label{CO sinrk}
\end{equation}\par 
The corresponding achievable rates of $s_c$ and $s_k$ at the $k$th user are $R_{c,k}({\bf P})=\log_2(1+\gamma_{c,k}({\bf P}))$ and $R_{k}({\bf P})=\log_2(1+\gamma_{k}({\bf P}))$. The common stream $s_c$ is decoded by all users. To ensure that all the $K$ users can successfully decode the common stream $s_c$, the corresponding rate should not exceed
\begin{equation}
R_c({\bf P})=\min \{R_{c,1}({\bf P}),\dots,R_{c,K}({\bf P})\}.
\end{equation}\par 
As $R_c({\bf P})$ is shared by $K$ users, we have $\sum_{k\in \mathcal{K}}C_k=R_c({\bf P})$, where $C_k$ is the portion of common rate at user-$k$ transmitting $W_{c,k}$. The total achievable rate of user-$k$ contains the portion of common rate transmitting $W_{c,k}$ and the private rate transmitting $W_{p,k}$. \par 
In this work, we aim at designing communication precoder $\bf P$ and message splits to maximize the WSR of downlink users and approximate the desired radar beampattern. Denote the weight allocated to user-$k$ as $\mu_k$, the formulated optimization problem can be written as 
\begin{subequations}\label{CO origin}
\begin{align}
\max\limits_{\alpha,{\bf c},{\bf P}} \quad &\sum_{k\in \mathcal{K}}\mu_k\left(C_k+R_k({\bf P})\right)\notag\\&-\lambda\sum_{m=1}^{M}\left|\alpha P_d(\theta_m)-{\bf a}^H(\theta_m)\left({\bf PP}^H\right){\bf a}(\theta_m)\right|^2\label{CO origin c3} \\ 
s.t.\quad&\sum_{k'\in \mathcal{K}}C_{k'}\leq R_{c,k}({\bf P}),\forall k\in\mathcal{K}\label{CO origin c1}\\
&{\bf c}\geq 0\label{CO origin c2}\\
&\text{diag}({\bf PP}^H)=\frac{P_t{\bf 1}}{N_t}\label{CO origin c4}\\
&\alpha>0\label{CO origin c5}
\end{align}
\end{subequations}
where $P_t$ is the total transmit power budget of the whole RadCom system, ${\bf c}=[C_1,\dots,C_K]^T$ is the common rate vector, the first part in \eqref{CO origin c3} maximizes the WSR from a communication perspective, while the second part ensures the probing beampattern approximates the desired pattern from a radar perspective, $\lambda$ is the regularization parameter that balances the communication WSR and radar beampattern approximation, \eqref{CO origin c1} ensures each user can decode the common stream in RS, \eqref{CO origin c4} ensures the transmit average power of each antenna to be the same. \par 
To show the advantage of RSMA in joint RadCom transmission, we consider the conventional multi-access SDMA based on multi-user linear precoding (MU-LP) in our RadCom system as a baseline. According to \cite{mao2018rate}, the SDMA-based RadCom design is obtained by allocating zero power to the common stream of RSMA-based RadCom, which is formulated by replacing the first part of \eqref{CO origin c3} with $\sum_{k\in \mathcal{K}}\mu_k R_k({\bf P})$ and removing \eqref{CO origin c1}. 
\section{ADMM-based Method for Solving the Problem }\label{admmdetail}
Since \eqref{CO origin} combines both communication and radar metrics, it is intuitive to think about alternately solving the communication and radar counterparts to find the optimal solution. In this section, we propose an iterative method based on ADMM to solve the nonconvex problem \eqref{CO origin}.\par 
To give an explicit expression of the approach, we first denote a new vector that contains all variables ${\bf v}=[\alpha, {\bf c}^T, {\text{vec}}({\bf P})^T]^T\in \mathbb{R}_{++}\times \mathbb{R}_+^{K}\times\mathbb{C}^{N_t\times(K+1)}$, with ${\bf p}_{\text{vec}}=\text{vec}({\bf P})$. 
We then define
\begin{equation}
\begin{split}
&{\bf F}_k=\text{Diag}(\underbrace{0,\dots,0}_{k},1,\underbrace{0,\dots,0}_{(K+1)N_t+K-k}), k\in\mathcal{K}\\
&{\bf D}_p=\text{Diag}(\underbrace{0,\dots,0}_{K+1},\underbrace{1,\dots,1}_{(K+1)N_t})\\
&{\bf D}_c=\text{Diag}(\underbrace{0,\dots,0}_{K+1},\underbrace{1,\dots,1}_{N_t},\underbrace{0,\dots,0}_{KN_t})\\
&{\bf D}_k=\text{Diag}(\underbrace{0,\dots,0}_{K+1+kN_t},\underbrace{1,\dots,1}_{N_t},\underbrace{0,\dots,0}_{(K-k)N_t}), k\in\mathcal{K}
\end{split}
\end{equation}
where $\text{Diag}(\cdot)$ is the diagonal matrix built from the entries within the bracket. We then further denote 
\begin{equation}
\begin{split}
&R_{c,k}({\bf P})=\eta_{c,k}({\bf p}_{\text{vec}})=\eta_{c,k}({\bf D}_p{\bf v}),\\
&R_{k}({\bf P})=\eta_{k}({\bf p}_{\text{vec}})=\eta_{k}({\bf D}_p{\bf v}),
\end{split}
\end{equation}
and rewrite \eqref{CO origin} in a tractable ADMM formulation
\begin{equation}
\begin{split}
\min\limits_{{\bf v},{\bf u}} \quad &f_c({\bf v})+g_c({\bf v})+f_r({\bf u})+g_r({\bf u})\\ 
s.t.\quad &{\bf D}_p({\bf v}-{\bf u})=0
\end{split} \label{CO admm}
\end{equation}
where ${\bf u}\in \mathbb{R}_{++}\times \mathbb{R}_+^{K}\times\mathbb{C}^{N_t\times(K+1)}$ is introduced as a new variable, 
\begin{equation}
f_c({\bf v})=-\sum_{k\in \mathcal{K}}\mu_k\left({\bf F}_k{\bf v}+\eta_k({\bf D}_p{\bf v})\right),  
\end{equation} 
 \begin{equation}
 \begin{split}
 f_r({\bf u})=&\lambda\sum_{m=1}^{M}|\alpha P_d(\theta_m)-{\bf a}^H(\theta_m)\big({\bf D}_c{\bf u}{{\bf u}}^H{\bf D}_c^H+\\
 &\sum_{k\in \mathcal{K}}{\bf D}_k{\bf u}{{\bf u}}^H{\bf D}_k^H \big) {\bf a}(\theta_m)| ^2,
 \end{split}
 \end{equation}
$g_c({\bf v})$ is the indicator function of the communication feasible set $\mathcal{C}=\left\{{\bf v}\bigg|\sum_{k\in \mathcal{K}}{\bf F}_k{\bf v}\leq \eta_{c,k}({\bf D}_p{\bf v})\right\}$,
 $g_r({\bf u})$ is the indicator function of radar feasible set $
 \mathcal{R}=\left\{{\bf v}\bigg|\text{diag}({\bf D}_c{\bf u}{{\bf u}}^H{\bf D}_c^H+\sum_{k\in \mathcal{K}}{\bf D}_k{\bf u}{{\bf u}}^H{\bf D}_k^H)=\frac{P_t{\bf 1}}{N_t}\right\}$.\par 
\eqref{CO admm} can be solved by iterating the following updates
\begin{align}
{\bf v}_r^{t+1}:=&\arg\min\limits_{{\bf v}_r}\big(f_c({\bf v}_r)+g_c({\bf v}_r)\notag\\
&+(\rho/2)\lVert {\bf D}_{pr}({\bf v}_r-{\bf u}_r^t)+{\bf d}_r^t\rVert_2^2\big)\label{updatev}\\
{\bf u}_r^{t+1}:=&\arg\min\limits_{{\bf u}_r}\big(f_r({\bf u}_r)+g_r({\bf u}_r)\notag\\
&+(\rho/2)\lVert{\bf D}_{pr}({\bf v}_r^{t+1}-{\bf u}_r)+{\bf d}_r^t\rVert_2^2\big)\label{updateu}\\
{\bf d}_r^{t+1}:=&{\bf d}_r^t+ {\bf D}_{pr}({\bf v}_r^{t+1}-{\bf u}_r^{t+1})\label{updated}
\end{align}
where ${\bf d}_r=[\mathfrak{R}\{{\bf d}\};\mathfrak{I}\{{\bf d}\}]^T$ with ${\bf d}\in \mathbb{C}^{N_t\times(K+1)}$ as the scaled dual variable. Here we write the ADMM expression in a real-valued way according to \cite{ma2010semidefinite} to avoid confusions, letting ${\bf v}_r=[\mathfrak{R}\{{\bf v}\};\mathfrak{I}\{{\bf v}\}]^T$, ${\bf u}_r=[\mathfrak{R}\{{\bf u}\};\mathfrak{I}\{{\bf u}\}]^T$, ${\bf D}_{pr}=\left[\mathfrak{R}\{{\bf D}_{p}\}, -\mathfrak{I}\{{\bf D}_{p}\}; \mathfrak{I}\{{\bf D}_{p}\},\mathfrak{R}\{{\bf D}_{p}\}\right]$, where $\mathfrak{R}\{\cdot\}$, $\mathfrak{I}\{\cdot\}$ respectively mean extracting real and imaginary parts. It's worth noting that this is just a different expression for the same problem to rigorously meet the definition of ADMM \cite{MAL-016}.\par 
Then we present how to solve \eqref{updatev} and \eqref{updateu} in each ADMM iteration.The ${\bf v}$-update  \eqref{updatev} is first equivalently rewritten in a tractable manner
\begin{equation}
\begin{split}\label{updatevorigin}
\min\limits_{{\bf c},{\bf P}} \quad &-\sum_{k\in \mathcal{K}}\mu_k\left(C_k+\eta_k({\bf p}_{\text{vec}})\right)+\frac{\rho}{2}\lVert {\bf p}_{\text{vec}}-{\bf D}_p{\bf u}^t+{\bf d}^t\rVert_2^2\\ 
s.t.\quad&\sum_{k'\in \mathcal{K}}C_k'\leq \eta_{c,k}({\bf p}_{\text{vec}}),\forall k\in\mathcal{K}\\
&{\bf c}\geq 0.\\
\end{split}
\end{equation}
This problem can be reformulated with Weighted Minimized Mean Square Errors (WMMSE) approach and solved through the WMMSE-based Alternating Optimization (AO) algorithm following \cite{mao2019RSWIPT}.\par 
The ${\bf u}$-update \eqref{updateu} is fully formulated as
\begin{equation}
\begin{split}\label{updateuorigin}
\min_{\alpha_u,{\bf p}_u}\quad&\lambda\sum_{m=1}^{M}\lvert P_d(\theta_m)\alpha_{u}-{\bf a}^H(\theta_m)\big(\sum_{k=1}^{K+1}{\bf D}_{p,k}{\bf p}_u{{\bf p}_u}^H{\bf D}_{p,k}^H\big) \\
&{\bf a}(\theta_m)\lvert ^2+\frac{\rho}{2}\lVert{\bf D}_p{\bf v}^{t+1}-{\bf p}_u+{\bf d}^t\rVert_2^2\\
s.t. \quad &\text{diag}(\sum_{k=1}^{K+1}{\bf D}_{p,k}{\bf p}_u{{\bf p}_u}^H{\bf D}_{p,k}^H)=\frac{P_t{\bf 1}}{N_t}\\
& \alpha_u\ge 0.
\end{split}
\end{equation} 
Here $\alpha_u=u_1$, ${\bf p}_u=\left[u_{K+2},u_{K+3},\dots,u_{(N_t+1)\times (K+1)}\right]^T$, where $u_i$ is the $i$th element in ${\bf u}$, and the selection matrix is defined as 
\begin{equation}
{\bf D}_{p,k}=\text{Diag}(\underbrace{0,\dots,0}_{(k-1)\times N_t},\underbrace{1,\dots,1}_{N_t},\underbrace{0,\dots,0}_{(K+1-k)\times N_t})
\end{equation}

\eqref{updateuorigin} is apparently non-convex, but can be solved by general Semidefinite Relaxation (SDR) method referring to \cite{ma2010semidefinite,boyd1994linear}. We summarize the ADMM-based method to solve problem \eqref{CO origin} in Algorithm 1, where ${\bf r}^{t+1}$ and ${\bf q}^{t+1}$ are the primal and dual residuals. Similarly, the SDMA-based RadCom problem can be solved via the proposed ADMM-based algorithm in this section. In the simulations, we see that the proposed solving method always converges within tens of iterations.
\begin{algorithm}\label{ADMMAL}
	\caption{ADMM-based method}
	\LinesNumbered 
	\KwIn{$t\leftarrow0,{\bf v}_r^{t},{\bf u}_r^{t},{\bf d}_r^{t}$}
	\Repeat{$\lVert{\bf r}^{t+1}\lVert_2\le \epsilon$ and $\lVert{\bf q}^{t+1}\lVert_2\le \epsilon$}{
Update ${\bf v}_r^{t+1}\gets \arg\min\limits_{{\bf v}_r}\big(f_c({\bf v}_r)+g_c({\bf v}_r)+(\rho/2)\lVert {\bf D}_{pr}({\bf v}_r-{\bf u}_r^{t})+{\bf d}_r^{t}\rVert_2^2\big)$ via WMMSE-based AO algorithm;\\
	Update ${\bf u}_r^{t+1}\gets \arg\min\limits_{{\bf u}_r}\big(f_r({\bf u}_r)+g_r({\bf u}_r)+(\rho/2)\lVert{\bf D}_{pr}({\bf v}_r^{t+1}-{\bf u}_r)+{\bf d}_r^{t}\rVert_2^2\big)$ via SDR-based algorithm;\\
	Update ${\bf d}_r^{t+1}\gets {\bf d}_r^k+ {\bf D}_{pr}({\bf v}_r^{t+1}-{\bf u}_r^{t+1})$;\\
	${\bf r}^{t+1}={\bf D}_{pr}({\bf v}_r^{t+1}-{\bf u}_r^{t+1})$;\\
	${\bf q}^{t+1}={\bf D}_{pr}({\bf u}_r^{t+1}-{\bf u}_r^{t})$;\\
	k++;
	}
\end{algorithm}

\section{Numerical Results}
\begin{figure}[htp]
	\begin{minipage}{\linewidth}
		\centering
		\includegraphics[width=\linewidth]{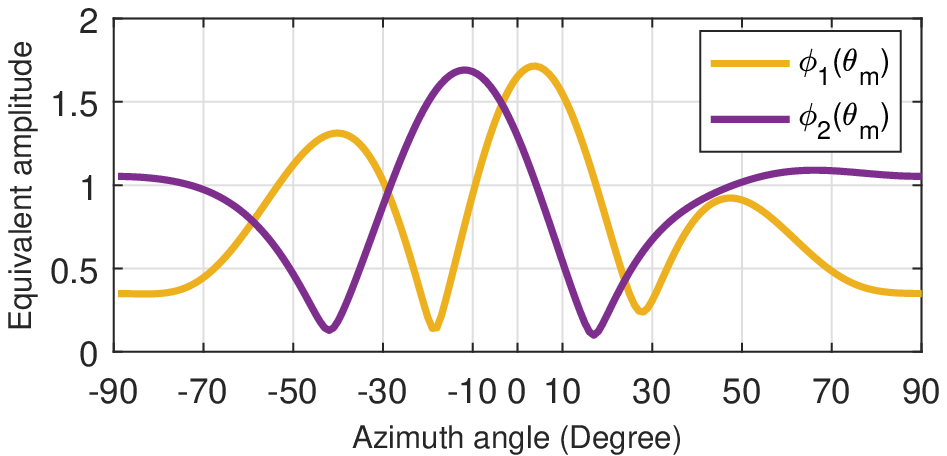}
		\caption{Equivalent amplitude of channel vectors.}\label{heffect}
	\end{minipage}	
	\begin{minipage}{\linewidth}
		\centering
		\includegraphics[width=\linewidth]{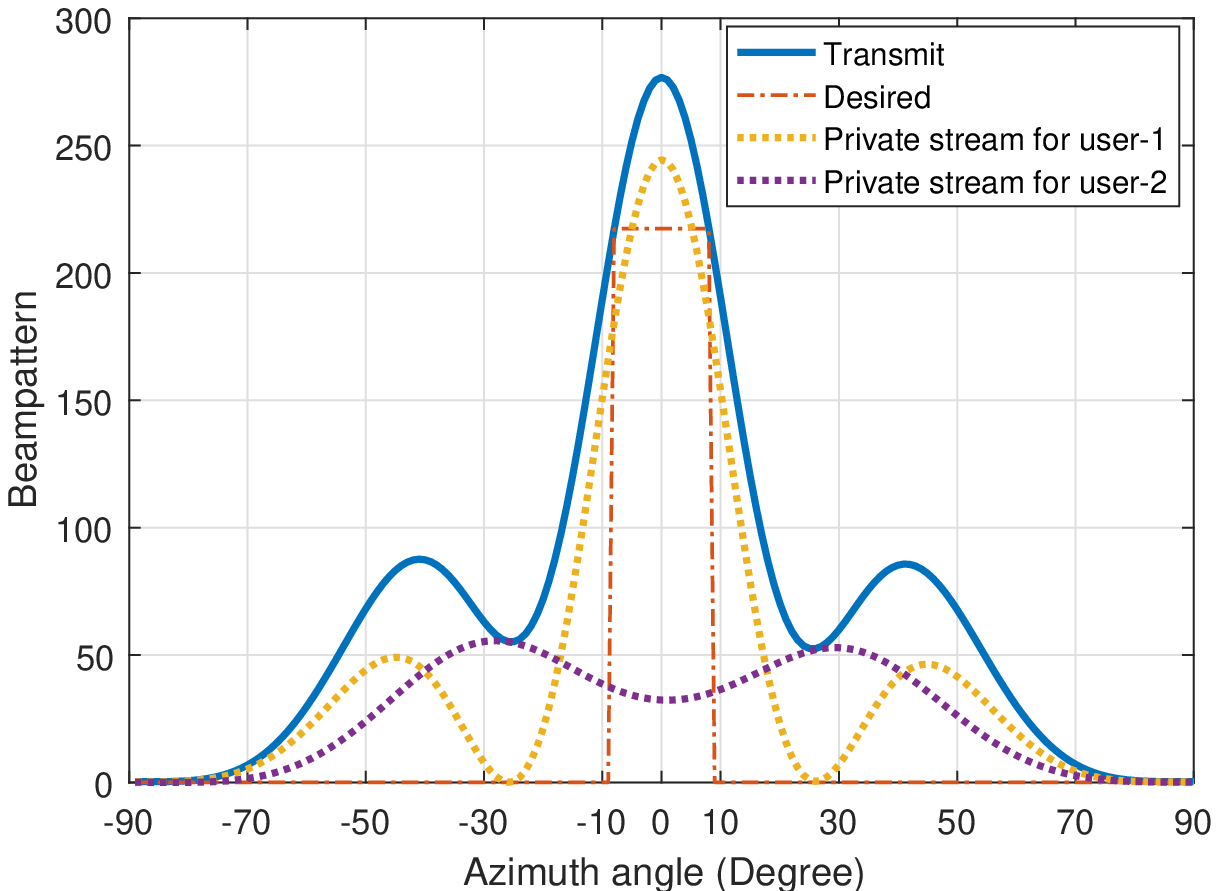}
		\caption{Transmit beampattern of SDMA-RadCom.}\label{MULP-beam}
	\end{minipage}
	\begin{minipage}{\linewidth}
		\centering
		\includegraphics[width=\linewidth]{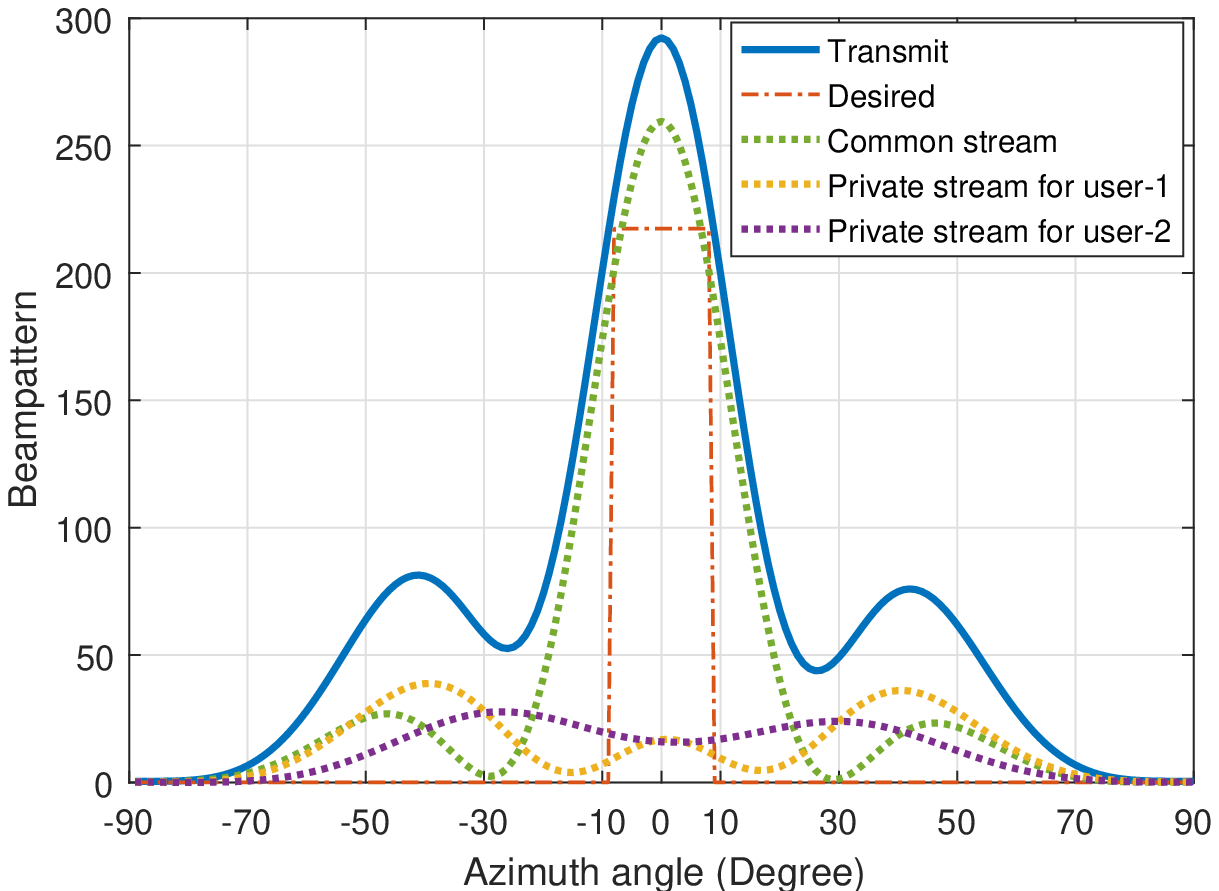}
		\caption{Transmit beampattern of RSMA-RadCom.}\label{RS-beam}
	\end{minipage}
\begin{minipage}{\linewidth}
\centering
\includegraphics[width=\linewidth]{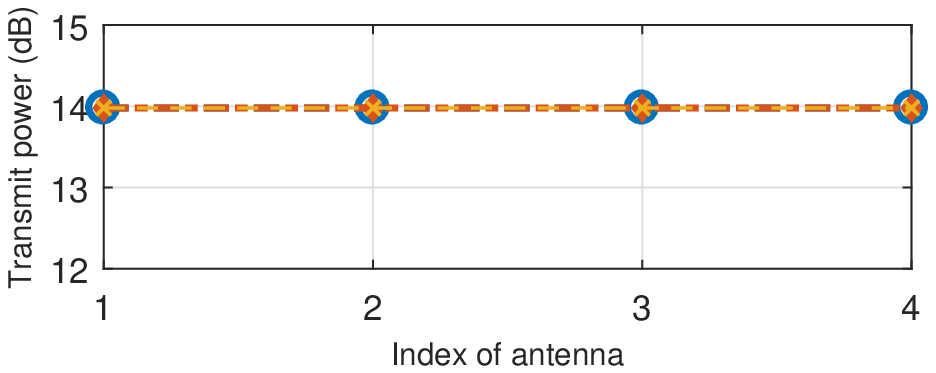}
\caption{Average power of each antenna.}\label{antennapower}
\end{minipage}
\end{figure}

In this part, numerical results are provided to validate the performance of the proposed RadCom transmission design. We assume $P_t=20\text{dBm}$, the noise power at each user is $0\text{dBm}$. We consider a simple case in this paper, where $K=2$, $N_t=4$,  $\mu_k=1/K,k\in\mathcal{K}$. The RadCom system employs a ULA with half-wavelength spacing, i.e., $\delta=0.5$. We randomly generate a pair of channel vectors obeying the i.i.d. standard complex Gaussian distribution, and use the specific channels in all the following experiments. The azimuth angle of the radar target of interest is 0\textdegree. We initialize ${\bf d}$ randomly obeying standard complex Gaussian distribution, ${\bf v}={\bf u}=[\alpha, {\bf c}^T, {\text{vec}}({\bf P})^T]^T$, where $\alpha=1$, ${\bf c}={\bf 1}^{K\times 1}$, and ${\bf P}$ is designed through Maximum Ratio Combining 
(MRC) method following \cite{joudeh2016sum}. Then the initialization ${\bf v}_r$, ${\bf d}_r$ and ${\bf u}_r$ can be obtained according to Section \ref{admmdetail}. The stopping criterion in Algorithm 1 is $\epsilon=10^{-2}$. We denote the jointly designed RadCom transmission based on RSMA and SDMA strategies respectively as RSMA-RadCom and SDMA-RadCom. We set the same channel vectors, desired beampattern and regularization parameter $\lambda=10^{-3}$ in Fig. \ref{heffect}-Fig. \ref{RS-beam} to compare both methods explicitly.  \par 
In Fig. \ref{heffect}, we first show the equivalent amplitude of each channel vector corresponding to each steering vector at $\theta_m$ in the beampattern. The amplitude is defined as $\phi_{k}(\theta_m)=|{\bf h}_k^H{\bf a}(\theta_m)|/(\lVert{\bf h}_k\lVert_2\cdot\lVert{\bf a}(\theta_m)\lVert_2)$. Fig. \ref{heffect} equivalently reflects each user's desired beampattern. It demonstrates that both users desire relatively high power near 0\textdegree, where the radar also desires a power peak. \par
We then compare the transmit beampattern of RSMA-RadCom and SDMA-RadCom in Fig. \ref{MULP-beam} and Fig. \ref{RS-beam}. The beampatterns of the precoded common stream and private streams are respectively displayed so as to show the contributions of all streams to the RadCom transmit beampattern. We can see from Fig. \ref{MULP-beam} that SDMA-RadCom forces the private stream to user-1 to contribute more to the radar's desired beam at 0\textdegree, as user-1 desires more power than user-2 at this angle. Although this strategy meets the radar's requirement, the high power of user-1's stream at 0\textdegree\  inevitably causes strong interference to user-2. However, from Fig. \ref{RS-beam}, we can see that the common stream instead takes the task of forming a beam at 0\textdegree. This is reasonable as the common stream benefits both users at 0\textdegree, which avoids the strong interference imposed by one user's stream upon the other in SDMA-RadCom. This indicates that introducing the common stream better mitigates the interference between users caused by the radar's beampattern requirement, which also explains why the beam peak at 0\textdegree\  and WSR of RSMA-RadCom are both higher than SDMA-RadCom.\par 
Fig. \ref{antennapower} shows the average power of each antenna, which verifies that the ADMM-based method reaches a high-quality solution to meet the average power constraint \eqref{CO origin c4} with a full use of the power budget $P_t$. Table \ref{Rateanalysis} analyses the rates achieved by RSMA-RadCom and SDMA-RadCom in this scenario. It displays that RSMA-RadCom has higher WSR and the common stream makes considerable contributions. \par
\begin{figure}
\centering
\includegraphics[width=\linewidth]{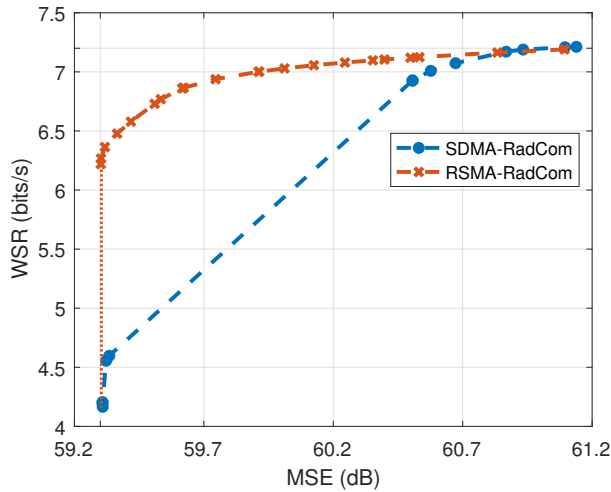}
\caption{Tradeoff comparison between WSR and MSE of beampattern approximation.}\label{tradeoff}
\end{figure}

 
 \begin{table}[htp]
 	\centering
 	\caption{Rate analysis for the senario in Fig. \ref{RS-beam} and Fig. \ref{MULP-beam}}\label{Rateanalysis}
 	\begin{tabular}[width=\linewidth]{cccccc}
 		\hline 
 		& ${\bf R}_1$ & $C_1$ & ${\bf R}_2$ & $C_2$ & WSR \\ 
 		\hline 
 		RSMA- & \multirow{2}[2]{*}{5.2894} &\multirow{2}[2]{*}{1.3454}  & \multirow{2}[2]{*}{3.9332} & \multirow{2}[2]{*}{1.3454} & \multirow{2}[2]{*}{5.9567} \\ 
 		RadCom& & & & &\\
 		\hline 
 		SDMA- & \multirow{2}[2]{*}{7.9120} &\multirow{2}[2]{*}{-}  & \multirow{2}[2]{*}{0.5098} & \multirow{2}[2]{*}{-} & \multirow{2}[2]{*}{4.2109} \\ 
 		RadCom& & & & &\\
 		\hline
 	\end{tabular} 
 \end{table} 
 Fig. \ref{tradeoff} is the tradeoff achieved by both RSMA-RadCom and SDMA-RadCom, which is obtained by varying $\lambda$ while keeping other settings the same. Obviously, Fig. \ref{tradeoff} further shows that RSMA-RadCom achieves a considerably better tradeoff than SDMA-RadCom in this scenario. \par     
\section{Conclusion}
To conclude, we propose a multi-antenna RSMA-RadCom system that functions both as a BS communicating with users and a radar detecting targets in specific azimuth angles of interest. We are the first to study maximizing WSR and approximating desired radar beampattern under the average power constraint of each antenna in the multi-antenna RadCom system. An ADMM-based method with WMMSE-based AO algorithm and SDR algorithm is proposed to solve the non-convex problem. Numerical results show that RSMA-RadCom achieves a better tradeoff compared with SDMA-RadCom in a specific scenario. This results from the creation of the common stream, which effectively mitigates interference especially at the transmit beampattern angles where we desire peaks. We can thus conclude that RSMA is a more powerful strategy when applied to multi-antenna RadCom design. However, different channel vectors may lead to different radar-communication tradeoff, which means further research is needed to fully evaluate the advantage of RMSA applied in RadCom in a variety of channel conditions.
\ifCLASSOPTIONcaptionsoff
  \newpage
\fi



\bibliographystyle{IEEEtran}
\bibliography{IEEEabrv,xcc}
\end{document}